\pdfoutput=1
\documentclass[submission,copyright,creativecommons]{eptcs}
\providecommand{\event}{MARS@ETAPS 2026} 

\usepackage{float}
\usepackage{amssymb}
\usepackage{iftex}
\usepackage{graphicx}
\usepackage{amsmath}
\usepackage{booktabs}
\usepackage{tikz}
\usetikzlibrary{automata,positioning,arrows.meta}
\ifpdf
  \usepackage{underscore}         
  \usepackage[T1]{fontenc}        
\else
  \usepackage{breakurl}           
\fi

\usepackage{xcolor}

\newcommand{\mv}{{MultiVeStA}}

\usepackage{listings}
\title{
Statistical Model Checking of the Island Model: An Established Economic Agent-Based Model of Endogenous Growth
	}
\author{Stefano Blando, Giorgio Fagiolo, Daniele Giachini, Andrea Vandin*
\institute{Institute of Economics and L'EMbeDS\\ Sant'Anna School of Advanced Studies\\Pisa, Italy}
\email{n.surname@santannapisa.it}
\and
Ernest Ivanaj 
\institute{Swiss Finance Institute\\ University of Geneve\\Switzerland}
\email{ernest.ivanaj@etu.unige.ch}
}

\def\titlerunning{Statistical Model Checking of the Island Model}
\def\authorrunning{S. Blando et al.}
\begin{document}
\maketitle

\begin{abstract}
Agent-based models (ABMs) are increasingly used to study complex economic phenomena such as endogenous growth, but their analysis typically relies on ad-hoc Monte Carlo exercises without formal statistical guarantees. We show how statistical model checking (SMC), and in particular MultiVeStA, can automate and enrich the analysis of a seminal ABM: the Island Model of Fagiolo and Dosi, which captures the exploration-exploitation trade-off in technological search. We reproduce key stylized facts from the original model with formal confidence intervals, confirm the optimality of moderate exploration rates, and perform a counterfactual sensitivity analysis across returns to scale, skill transfer, and knowledge locality. Using MultiVeStA's built-in Welch's t-test, 6 out of 7 pairwise parameter comparisons yield statistically different growth trajectories, while the exception reveals a saturation effect in knowledge locality. Our results demonstrate that SMC offers a principled, reproducible methodology for the quantitative analysis of agent-based economic models.

\end{abstract}

\section{Introduction}
\label{sec:intro}


One of the key challenges in economics is understanding how economic systems evolve over time, and in particular identifying the sources of long-run economic growth. Traditional growth models, such as the Solow-Swan model~\cite{solow1956contribution}, treat technological progress as exogenous, leaving unexplained the very engine of sustained growth. \emph{Endogenous growth theory} emerged as a response, with contributions by Romer~\cite{romer1990endogenous}, Lucas~\cite{lucas1988mechanics}, and Aghion and Howitt~\cite{aghion1992model}, explaining growth through internal mechanisms such as innovation, human capital, and knowledge spillovers (see Section~\ref{sec:growth}). However, these models usually rely on simplifying assumptions such as representative agents, rational behavior, and equilibrium, which limit their adherence to real economic dynamics characterized by heterogeneity and interaction at the micro level~\cite{kirman2011complex}.

More recently, agent-based computational economics (ACE) has been proposed as a solution to these problems. ACE offers a complementary approach~\cite{tesfatsion2006handbook,dawid2018agent} where economic dynamics are modeled as collections of simple interacting agents with heterogeneous characteristics and adaptive behaviors, i.e. agent-based models (ABMs). These models, also known as multi-agent systems~\cite{wooldridge2009introduction}, can capture emergent macroeconomic phenomena such as growth, business cycles, or crises from interaction at the individual level (see Section~\ref{sec:abm}). However, ABMs do not admit analytical solutions and their analysis relies on numerical simulations, where coming up with the \emph{right} experiment design is difficult but crucial to obtain meaningful insights~\cite{Secchi2017,Vandin2022}.

In a recent line of research~\cite{Vandin2022,DBLP:conf/vecos/Vandin24,DBLP:journals/corr/abs-2509-10977}, approaches from computer science known as \emph{statistical model checking} (SMC)~\cite{legay2010statistical,agha2018survey} have been proposed as a solution to the problem of rigorous analysis of ABMs. This research line builds on MultiVeStA~\cite{sebastio2013multivesta,GilmoreRV17,Vandin2022}, a statistical model checker designed for the quantitative analysis of discrete-event simulations. Compared to standard Monte Carlo approaches, SMC automatically determines the minimum number of simulations needed to achieve user-specified confidence levels, provides built-in formal hypothesis testing to compare different model configurations, and can target temporal properties through expressive query languages.

In this paper, we show how SMC, and in particular \mv{}, can be used to automate the analysis of a seminal ABM, the \emph{Island Model} of Fagiolo and Dosi~\cite{fagiolo2003exploitation}. This model captures the exploration-exploitation tradeoff in technological search.
Our goal is methodological rather than model-theoretic: we treat the Island Model as a black-box stochastic transition system observed through aggregate outputs, and show how \mv{} can equip its analysis with formal statistical guarantees---without re-encoding it in a formal modelling language.
Our contributions are: (1) a detailed description of the Island Model accessible to non-experts; (2) software engineering aspects connected to the integration of the model with \mv{};
(3) a discussion of the obtained results, including a counterfactual analysis with formal pairwise hypothesis testing across different system parametrizations.

The paper is organized as follows. Section~\ref{sec:abm} provides background on agent-based modeling. Section~\ref{sec:growth} reviews endogenous growth theory. Section~\ref{sec:smc} introduces statistical model checking and MultiVeStA. Section~\ref{sec:island} presents the Island Model. Section~\ref{sec:integration} describes our integration. Section~\ref{sec:analysis} presents results. Section~\ref{sec:conclusions} concludes.

\section{Agent-Based Modeling in Economics}
\label{sec:abm}
Agent-based computational economics (ACE) models economies as collections of interacting autonomous agents~\cite{tesfatsion2006handbook}. Unlike traditional economic modeling based on representative agents and equilibrium analysis, ACE studies how macroeconomic patterns emerge from microeconomic interactions of heterogeneous individuals with bounded rationality.

\paragraph{Foundations.}
The foundations of ACE draw from multiple sources. Herbert Simon's work on bounded rationality~\cite{simon1955behavioral} challenged perfect optimization assumptions, arguing that real decision-makers use heuristics and satisficing strategies. The Santa Fe Institute's artificial stock market~\cite{arthur1996asset} demonstrated how complex market dynamics emerge from adaptive trading rules. Thomas Schelling's segregation model~\cite{schelling1971dynamic} showed how aggregate patterns arise from mild individual preferences.

Several characteristics distinguish ACE from traditional macroeconomic modeling~\cite{dawid2018agent}:
    \emph{Heterogeneity}, Agents differ in preferences, beliefs, strategies, and constraints, contrasting with representative-agent models; 
    \emph{Bounded rationality}, Agents use heuristics and learning rules rather than solving optimization problems; 
    \emph{Local interactions}, Agents interact with subsets of others rather than through centralized markets; 
    \emph{Out-of-equilibrium dynamics}, ABMs model transition paths and crises, not just steady states; 
    \emph{Emergence},  Macroeconomic phenomena arise from microeconomic interactions without being programmed.

\paragraph{Applications and Challenges.}
ABMs have been applied across economics: financial markets and bubbles~\cite{bottazzi2005institutional,hommes2006heterogeneous,lux2009economics,anufriev2012asset,bottazzi2017wealth}, systemic risk~\cite{battiston2016complexity}, macroeconomic policy~\cite{dawid2012eurace}, and innovation dynamics~\cite{nelson1982evolutionary,dosi2010schumpeter,Fagiolo2020InnovationFinanceGrowth, capone2019history}.
The Island Model belongs to this tradition of studying growth through technological search.

Despite flexibility, ABMs pose analytical challenges~\cite{fabretti2013problem,windrum2007empirical}. Stochasticity means outcomes are distributions requiring many runs. Parameter sensitivity demands systematic exploration. Validation against empirical data is difficult. Statistical model checking addresses the first two challenges through rigorous, automated methods with quantifiable confidence.

\section{Endogenous Growth Theory}
\label{sec:growth}

Endogenous growth theory explains long-run growth from within the economic system, rather than treating technological progress as exogenous~\cite{romer1994origins}.

\paragraph{From Exogenous to Endogenous Growth.}
The Solow-Swan model~\cite{solow1956contribution} demonstrated that capital accumulation alone cannot sustain growth due to diminishing returns. The economy converges to a steady state where output grows only with exogenous technological progress. This explains convergence but leaves growth's ultimate source unexplained.

Endogenous growth theory filled this gap. The key insight was that knowledge differs from physical capital: ideas are non-rival (one person's use doesn't diminish another's) and potentially non-excludable. These properties create increasing returns at aggregate level~\cite{romer1990endogenous}. Romer's model formalized how firms invest in R\&D to develop ideas that create spillovers benefiting others. Aghion and Howitt~\cite{aghion1992model} developed a Schumpeterian approach emphasizing creative destruction.


The key mechanisms in endogenous growth include: 
     \emph{Learning by doing}, productivity improves as byproduct of experience~\cite{arrow1962economic}; 
   \emph{Human capital}, education increases productivity and innovation capacity~\cite{lucas1988mechanics}; 
    \emph{R\&D}, intentional research generates new technologies~\cite{romer1990endogenous}; 
    \emph{Knowledge spillovers}, ideas spread between firms and regions~\cite{jaffe1993geographic}; 
   \emph{Increasing returns}, scale economies create positive feedback~\cite{krugman1991increasing}. 

\paragraph{The Evolutionary Tradition and the Exploration-Exploitation Tradeoff.}
A foundational contribution to the agent-based study of economic growth is the evolutionary theory of Nelson and Winter~\cite{nelson1982evolutionary}, which models firms as boundedly-rational entities that search for new technologies through stochastic routines rather than optimizing behavior. In their framework, innovation is an evolutionary process driven by variation (search for new techniques), selection (market competition), and retention (organizational routines). This perspective---where growth emerges from the collective dynamics of heterogeneous firms rather than from a representative agent's optimization---provides the intellectual foundation for the Island Model and, more broadly, for the Schumpeterian tradition in agent-based economics~\cite{dosi2010schumpeter,dosi2017micro}.

A central theme within this tradition is the exploration-exploitation tradeoff, formalized by March~\cite{march1991exploration} in the context of organizational learning. March showed that organizations face a fundamental tension: \emph{exploitation} of existing competencies yields reliable but incremental returns, while \emph{exploration} of new alternatives is uncertain but potentially transformative. Crucially, the two strategies compete for scarce resources, and an excess of either leads to suboptimal outcomes---pure exploitation causes lock-in and stagnation, while pure exploration dissipates resources on unproven alternatives. This insight, further developed by Levinthal~\cite{levinthal1993myopia} in terms of the ``myopia of learning,'' has been influential across organizational theory, evolutionary economics, and machine learning (multi-armed bandits)~\cite{sutton2018reinforcement}.

The Island Model~\cite{fagiolo2003exploitation} directly operationalizes March's tradeoff: agents choose between mining known islands (exploitation) and searching for new ones (exploration), with the aggregate balance determining the economy's growth trajectory. The model further connects to fitness landscape theory~\cite{kauffman1993origins} and the literature on technological search~\cite{fleming2001recombinant}, where the structure of the search space shapes the returns to exploration.

These ABMs of endogenous growth pose significant analytical challenges: their large state spaces and path-dependent dynamics preclude closed-form solutions, while the sensitivity of growth trajectories to parameter values demands systematic exploration backed by statistical guarantees. These features make them natural candidates for statistical model checking, which we introduce next.

\section{Statistical Model Checking and MultiVeStA}
\label{sec:smc}

The communities of verification and formal methods, part of computer science, proposed over the years several techniques for the analysis of systems. A notable example is the family of techniques known as model checking \cite{DBLP:books/daglib/0007403-2,DBLP:books/daglib/0020348}.

\paragraph{Model Checking.}
Traditional model checking verifies whether system $M$, given in some mathematical formalism, satisfies a property $\phi$, typically given in a formal logic, written $M \models \phi$ \cite{DBLP:books/daglib/0007403-2}. For finite state spaces, this can be done exhaustively. However, ABMs have state spaces too large or infinite for exhaustive checking.

Probabilistic versions of model checking (PMC)~\cite{DBLP:books/daglib/0020348,10.1145/2933575.2934574} can handle stochastic systems, but still require state space exploration.  Instead of ``does $M$ satisfy $\phi$?'', these answer questions like ``what is the probability that $M$ satisfies $\phi$?''.
Statistical model checking (SMC) \cite{agha2018survey,legay2010statistical} answers this question by estimating such probabilities. It simulates $M$ multiple times, evaluating $\phi$ on each trace, and using statistics to estimate the probability.

The key advantage is scalability: required simulations depend on desired confidence and precision, not state space size. This comes at the price of losing exactness in the analysis results. However, the statistical guarantees provided by SMC can be fine tuned.
Two approaches exist~\cite{agha2018survey}: hypothesis testing (using Sequential Probability Ratio Test~\cite{wald1945sequential}) and direct estimation with confidence intervals.

\paragraph{MultiVeStA.}

MultiVeStA~\cite{sebastio2013multivesta,Vandin2022,GilmoreRV17} is a statistical analyzer that performs quantitative analysis on models of systems, with the only requirement being that of admitting probabilistic/stochastic (IID) simulations. Its architecture supports black-box integration with existing discrete-event simulators through a simple API. This makes SMC more accessible to external domains, allowing to directly analyze existing models written in non-formal general-purpose programming languages, or agent-based domain-specific languages.  
For example, \mv{} has been applied to several domains and simulators such as crowd steering scenarios~\cite{DBLP:conf/hpcs/PianiniSV14}, security threat analysis~\cite{DBLP:journals/corr/abs-2101-08677,CasaluceBCV23},
public transportation systems in smart cities~\cite{DBLP:conf/ifm/GilmoreTV14,DBLP:conf/isola/CianciaLMPV16},  product lines engineering~\cite{DBLP:journals/tse/BeekLLV20,DBLP:conf/fm/VandinBLL18}, business processes~\cite{DBLP:journals/jss/CorradiniFPRTV21},
decentralized finance~\cite{DBLP:conf/isola/BartolettiCJLMV22}, robotic systems~\cite{DBLP:conf/birthday/BelznerNVW14,DBLP:journals/scp/BruniCGLV15}, collective adaptive systems \cite{DBLP:conf/wsc/GalpinGLV18,DBLP:conf/isola/CasaluceTV24}, agent-based models~\cite{Vandin2022,DBLP:conf/vecos/Vandin24, DBLP:journals/corr/abs-2509-10977}. This included java-, c- and python-based simulators. 

\mv{} alleviates the burden coming from performing reliable statistical analyses, as it automates all involved steps (e.g., triggering the minimum number of simulations required for obtaining user-specified confidence intervals, or comparing the results obtained for different model parameterizations).
In particular, \mv{} enables the verification of properties defined in the MultiQuaTEx language, a practitioner-oriented domain-specific language allowing recursive and parametric queries corresponding to formal logics. 
%


We complete this section with a brief introduction to the MultiQuaTEx language, which we use to express the properties of interest in our analysis. 
%
MultiQuaTEx uses recursive functions over simulation states. The basic building block is \texttt{s.rval("x")} returning observable \texttt{x}'s current value. For example, to compute expected log(GDP) at time 101:

\begin{lstlisting}
obsAtStep(x, obs) =
  if (s.rval("steps") == x) then s.rval(obs)
 else # obsAtStep(x,obs) fi;

\end{lstlisting}

The function recursively advances simulation (via \texttt{\#}) until reaching the target step. Parametric queries enable systematic analysis:

\begin{lstlisting}
eval parametric(E[ obsAtStep(t, "logGDP") ], t, 1, 10, 201);
\end{lstlisting}

For each simulation, this computes $\log(\text{GDP}(t))$ for $t \in \{1, 11, 21, \ldots, 201\}$. By performing enough simulations, \mv{} uses these values to estimate the expected values of $\log(\text{GDP})$ in each time point, each equipped with a confidence interval of width specified by the user.
Indeed, SMC suits ABMs well: the stochasticity and large state spaces of ABMs make exhaustive analysis particularly challenging, while SMC provides principled uncertainty quantification and systematic parameter exploration.
%
%
In particular, given a desired statistical significance level $\alpha$ and interval width $\delta$, \mv{} automatically determines and performs the minimum number of simulations needed to guarantee that each point estimate lies within a confidence interval of width at most $\delta$, with statistical confidence $1 - \alpha$~\cite{Vandin2022}.

\paragraph{Comparison with Alternative Approaches.}
The analysis of ABMs through simulation has a long tradition, and several methodologies exist beyond plain Monte Carlo. \emph{Traditional sensitivity analysis}~\cite{saltelli2008global} varies parameters one at a time (or via factorial designs) and records the effect on output statistics, but typically relies on a fixed, predetermined number of simulations without formal guarantees on the precision of the estimates. \emph{Parametric grid search} systematically covers the parameter space but faces the curse of dimensionality and, again, offers no principled criterion for the required sample size at each configuration. \emph{Econometric meta-modeling}~\cite{kleijnen2015design} fits surrogate models (e.g., regression or kriging) to simulation output, enabling efficient interpolation across the parameter space, but requires careful experimental design and may miss non-smooth or regime-switching behaviors common in ABMs. More recently, machine learning surrogates have been proposed to accelerate ABM calibration and parameter exploration~\cite{lamperti2018agent}, combining neural networks and gradient-boosted trees with intelligent sampling to efficiently navigate large parameter spaces. Comprehensive frameworks for the empirical validation of ABMs have also been developed~\cite{fagiolo2007critical,fagiolo2019validation}, addressing the interplay between calibration, sensitivity analysis, and comparison with empirical data.

SMC, as implemented in \mv{}, complements these approaches by offering \emph{adaptive} sample sizes that are automatically determined to achieve user-specified statistical guarantees at every point of interest. Rather than fixing the number of runs ex ante, \mv{} adds simulation batches until the confidence interval width falls below the target $\delta$, ensuring that regions of high variance receive proportionally more runs. Furthermore, \mv{}'s built-in hypothesis testing provides a rigorous framework for counterfactual comparisons that does not require fitting an intermediate model, operating directly on the simulation output with controlled Type~I error and computable statistical power~\cite{Vandin2022}.

\section{The Island Model: An ABM to Study Endogenous Growth}
\label{sec:island}






The Island Model~\cite{fagiolo2003exploitation} captures the exploration-exploitation tradeoff in technological search. Inspired by Phelps' islands economy~\cite{phelps1970microeconomic}, it reinterprets ``islands'' as technologies in an abstract technology space.

\subsection{Economic Motivation}

The model translates core mechanisms of endogenous growth into an agent-based framework through precise analogies. \emph{Exploration} corresponds to R\&D investment: agents who leave a productive island to search for new ones bear a direct opportunity cost (foregone output) in exchange for the chance of discovering a superior technology---mirroring how firms allocate resources between current production and speculative research. \emph{Imitation} captures technological diffusion: agents who observe a stronger productivity signal from a distant island migrate toward it, analogous to firms adopting proven innovations from competitors~\cite{nelson1982evolutionary}.

The spatial structure of the grid encodes the notion that more radical innovations---further from the known technological frontier (the center)---tend to be more productive but harder to reach, reflecting the empirical regularity that breakthrough technologies require longer search but yield higher returns~\cite{fleming2001recombinant}. The parameter $\phi$ governs \emph{cumulative learning}: past skills carry over to newly discovered islands, capturing the learning-by-doing mechanism of Arrow~\cite{arrow1962economic} whereby a firm's absorptive capacity grows with experience. The parameter $\rho$ controls the spatial decay of productivity signals, modeling knowledge spillovers: low $\rho$ corresponds to a regime of broad information diffusion (e.g., open science, strong patent disclosure), while high $\rho$ restricts information to local clusters. Finally, $\alpha$ determines returns to scale in production---whether concentrating workers on a single technology yields increasing ($\alpha > 1$) or decreasing ($\alpha < 1$) marginal returns---directly connecting to the debate on agglomeration economies~\cite{krugman1991increasing}.

\subsection{Model Components}

\textbf{Technology space}: A $T \times T$ grid where each cell $(x,y)$ may contain an island with probability $\pi$. The center $(T/2, T/2)$ always contains an island where all agents start, representing the initial technology.

\textbf{Agents}: $N$ agents occupy grid positions in one of three states:
\begin{enumerate}
    \item \textbf{Miners} (Type 1): Exploit known islands, producing output and broadcasting productivity signals
    \item \textbf{Imitators} (Type 2): Target successful islands, modeling the adoption of established technologies
    \item \textbf{Explorers} (Type 3): Search randomly for new islands, representing R\&D
\end{enumerate}

\textbf{Island productivity}: When discovered, the productivity of a new island/technology in position $(x,y)$ is determined by:
\begin{equation}
    s_{(x,y)} = (1 + \text{Poisson}(\lambda)) \cdot \left( |x - T/2| + |y - T/2| + \phi \cdot \text{skills}_i + \epsilon \right)
\end{equation}
Here, the Poisson term represents breakthroughs, distance from center captures the novelty premium (more distant technologies tend more productive), past skills reflect cumulative learning, and $\epsilon \sim N(0,1)$ adds noise.

\textbf{Production}: Miners at island $(x,y)$ produce:
\begin{equation}
    y_i = s_{(x,y)} \cdot m_{(x,y)}^{\alpha - 1}
\end{equation}
Here, $m_{(x,y)}$ is miner count and $\alpha$ controls returns to scale ($\alpha < 1$: decreasing returns/crowding; $\alpha = 1$: constant; $\alpha > 1$: increasing returns/agglomeration). GDP is total production across miners.

\subsection{Dynamics}

At each time step:

\textbf{Signal transmission}: Miners broadcast signals decaying with distance. Agent $i$ receives signal from miner $j$ with probability:
\begin{equation}
    w_{ij} = \frac{m_{(x_j, y_j)}}{\sum_k \mathbf{1}[\text{Type}_k = 1]} \cdot \exp(-\rho \cdot d_{ij})
\end{equation}
Here, $d_{ij}$ is Manhattan distance. Parameter $\rho$ controls knowledge locality: low $\rho$ creates global information; high $\rho$ creates local bubbles.

\begin{figure}[h]
\centering
\includegraphics[width=0.75\textwidth]{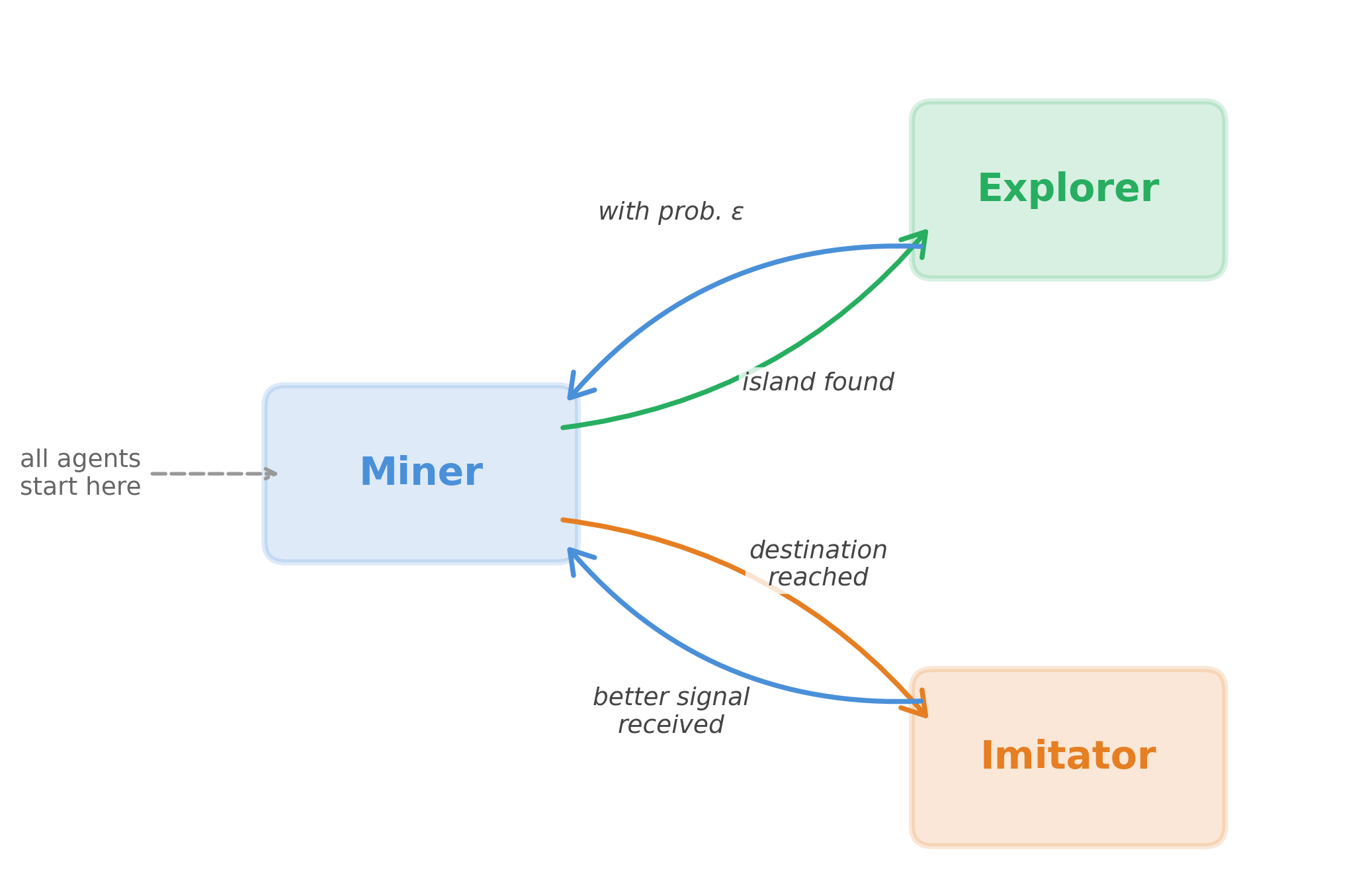}
\caption{Agent type transitions. Miners are the productive state. Exploration (with probability $\varepsilon$) and imitation (upon receiving a stronger productivity signal) represent two distinct search strategies. Both explorers and imitators return to mining upon completing their search.}
\label{fig:transitions}
\end{figure}

\textbf{Type transitions}: At each step, each miner may become an explorer with probability $\varepsilon$, representing the willingness to abandon a known technology in search of a better one. Alternatively, a miner who receives a productivity signal stronger than its current production becomes an imitator, heading toward the more productive island. Conversely, an explorer who lands on an undiscovered island becomes a miner, with the island's productivity determined at the moment of discovery. Similarly, an imitator who reaches its destination island reverts to mining. These transitions are illustrated in Fig.~\ref{fig:transitions}.

\textbf{Movement}: Explorers move randomly in cardinal directions; imitators move deterministically toward destinations.

Table~\ref{tab:params} summarizes parameters with economic interpretations.

\begin{table}[ht]
\centering
\caption{Island Model Parameters}
\label{tab:params}
\begin{tabular}{@{}clcc@{}}
\toprule
Parameter & Description & Default & Economic Interpretation \\
\midrule
$N$ & Number of agents & 20 & Labor force size \\
$T$ & Simulation length & 201 & Time horizon (steps) \\
$\pi$ & Island density & 0.1 & Technological opportunity \\
$\alpha$ & Returns to scale & 1.5 & Market structure \\
$\varepsilon$ & Exploration probability & 0.1 & Innovation intensity \\
$\lambda$ & Technology jump & 1 & Breakthrough frequency \\
$\phi$ & Past skills weight & 0.5 & Learning-by-doing strength \\
$\rho$ & Knowledge locality & 0.1 & Information regime \\
\bottomrule
\end{tabular}
\end{table}

\paragraph{Why Statistical Model Checking.}
The Island Model exhibits several features that make it particularly challenging to analyze with standard Monte Carlo methods. First, the dynamics are strongly \emph{path-dependent}: early exploration successes or failures can lock the economy into qualitatively different growth regimes, leading to high inter-simulation variance. Second, the model can produce \emph{multiple dynamic regimes}---sustained growth, stagnation, or lock-in on suboptimal technologies---depending on the stochastic sequence of discoveries and agent transitions. Third, the sensitivity to parameters such as $\alpha$, $\rho$, and $\varepsilon$ is non-trivial: small changes can shift the economy from one regime to another, but this is difficult to detect without controlled confidence intervals at each time point.

A standard Monte Carlo approach computes sample means over a fixed number of runs, but provides no formal guarantee on the precision of those estimates, nor built-in tools to rigorously compare different parametrizations. SMC, and in particular \mv{}, addresses these limitations by automatically determining the number of simulations required to achieve a target confidence interval width at \emph{every} time point of interest, and by providing built-in hypothesis testing to formally establish whether two parameter configurations produce statistically distinguishable trajectories.  Our analysis focuses on aggregate observables (GDP, logGDP, AGR), expected trajectories over a finite horizon ($T=201$), and one-parameter-at-a-time sweeps, choices that privilege interpretability and comparability with the original paper at the cost of leaving aside agent-level distributions and joint parameter effects
\paragraph{Original Implementation.}
The model was originally a monolithic MATLAB script combining parameters, state variables, and dynamics in a single file. While functional, this structure posed challenges: limited modularity (difficult to modify aspects independently), poor reusability (no stepping or intermediate queries), difficult external integration, and hard extensibility. These motivated the restructuring described next, which also required controlling random seeds for independent runs and preserving the order of random draws to maintain behavioral equivalence with the original.

\section{Model Implementation and Integration Details}
\label{sec:integration}

In this section, we present code-specific implementation details of the model, as well as aspects connected to its integration with \mv. The original monolithic script was refactored into a modular, object-oriented architecture exposing the interfaces required by the statistical model checker.


\subsection{Architecture Overview}

We restructured into two main classes:
\begin{itemize}
    \item \textbf{Model}: Main simulation class containing global state (grid, GDP, discoveries) and dynamics
    \item \textbf{Agent}: Individual agent class encapsulating state (position, type, production) and behaviors (movement, transitions)
\end{itemize}

This separation allows extending with new agent types or behaviors without modifying core logic.

\subsection{Model Interface}

Integrating a simulator with \mv{} requires exposing three basic actions~\cite{Vandin2022}: (i) \texttt{reset(seed)}, which resets the simulator to its initial state and updates the random seed used for pseudo-random number generation, so that each simulation run is independent; (ii) \texttt{next}, which advances the simulation by one step; (iii) \texttt{eval(obs)}, which evaluates an observation on the current simulation state, where an observation can be any feature of the aggregate model or of any group of agents. In our MATLAB implementation, these correspond to methods of the \texttt{Model} class:

\begin{lstlisting}[language=Matlab, caption={MultiVeStA integration interface (Model class).}, label={lst:mv-interface}]
classdef Model
  methods
    function obj = setParams(obj, pi, alpha, eps, phi, rho, lambda, ...)
    function obj = reset(obj, seed)           % action (i)
    function obj = next(obj)                  % action (ii)
    function value = evalObs(obj, variable)   % action (iii)
  end
end
\end{lstlisting}

An additional method, \texttt{setParams}, receives model parameters as strings from the command line and is called once at startup, enabling parameter sweeps via the \texttt{-otherParams} flag. The observable interface provides queryable quantities at each step:

\begin{lstlisting}[language=Matlab, caption={Observable interface.}, label={lst:evalobs}]
function value = evalObs(obj, variable)
    switch variable
        case "GDP"
            value = obj.GDP(obj.CurrentStep);
        case "logGDP"
            value = log(obj.GDP(obj.CurrentStep));
        case "AGR"
            value = obj.GDP(obj.CurrentStep) ...
                  - obj.GDP(obj.CurrentStep-1);
        case "AGR_total"
            gdp_start = obj.GDP(t_start);
            gdp_end   = obj.GDP(obj.CurrentStep);
            value = (log(gdp_end) - log(gdp_start)) ...
                  / (obj.CurrentStep - t_start + 1);
    end
end
\end{lstlisting}

Observables include \texttt{GDP} (raw aggregate output), \texttt{logGDP} (logarithm of aggregate GDP, used for growth analysis), \texttt{AGR} (absolute growth rate between consecutive steps), and \texttt{AGR\_total} (average growth rate over the entire simulation, computed as $(\log \text{GDP}_{T} - \log \text{GDP}_{t_0}) / (T - t_0)$, where $t_0$ is the first step with positive GDP).

The \texttt{-otherParams} flag specifies the class name, the evaluation method, the parameter method, and the parameter values ($\pi$, $\alpha$, $\varepsilon$, $\varphi$, $\rho$, $\lambda$). MultiVeStA then calls \texttt{setParams} once, and iterates \texttt{reset}/\texttt{next}/\texttt{evalObs} for each simulation run until convergence.

\clearpage
\subsection{Agent Class}

The Agent class encapsulates individual state and behavior:
\begin{lstlisting}[language=Matlab]
classdef Agent
    properties
        Type          % 1: Miner, 2: Imitator, 3: Explorer
        X; Y          % grid position
        Productivity; Production; Past_skills
    end
    methods
        function obj = Agent(type, x, y)
            obj.Type = type; obj.X = x; obj.Y = y;
            obj.Productivity = 1; obj.Production = 0; obj.Past_skills = 1;
        end
        function obj = move(obj, d)
            if d=="right", obj.X=obj.X+1; elseif d=="left", obj.X=obj.X-1;
            elseif d=="up", obj.Y=obj.Y+1; elseif d=="down", obj.Y=obj.Y-1; end
        end
\end{lstlisting}

Additional methods handle type transitions (\texttt{becomeMiner}, \texttt{becomeExplorer}, \texttt{becomeImitator}) and production.

This encapsulation enables adding new agent types or modifying behaviors without affecting the simulation loop.


\subsection{MultiQuaTEx Queries}

We developed two queries for different analyses:

\textbf{Transient analysis of log(GDP)} studies the evolution of output over time:
\begin{lstlisting}
obsAtStep(x, obs) =
  if (s.rval("my_time") == x) then s.rval(obs)
  else # obsAtStep(x, obs) fi;
eval parametric(E[ obsAtStep(x, "logGDP") ], x, 1, 10, 201);
\end{lstlisting}

This computes the average $\log(\text{GDP}(t))$ for $t \in \{1, 11, 21, \ldots, 201\}$, producing a time series of average log-output with confidence intervals. It is used for all parameter sweeps.

\textbf{Average Growth Rate (AGR)} computes the overall growth rate at the end of the simulation, used for the exploration-exploitation analysis:
\begin{lstlisting}
obsAtStep(x,obs) =
 if ( s.rval("my_time") == x )
  then s.rval(obs)
  else # obsAtStep(x,obs) fi ;
eval E[ obsAtStep(201,"AGR_total") ];
\end{lstlisting}

The execution pipeline consists of: configuration specification, query selection, MultiVeStA execution with block-based convergence checking, CSV output with means and confidence intervals, and post-processing visualization.

\section{Analysis}
\label{sec:analysis}
We now present the results of applying \mv{} to the Island Model. We first reproduce two stylized facts from the original paper, then perform a counterfactual sensitivity analysis with formal hypothesis testing across different parameter configurations.

\subsection{Experimental Setup}

All experiments use a 95\% confidence level ($\alpha_{\text{conf}} = 0.05$), block size of 30 simulations, and a precision threshold $\delta = 1$ for the width of the confidence interval. MultiVeStA automatically determines the number of simulation batches required to achieve convergence. The baseline configuration uses the defaults from Table~\ref{tab:params}. Each simulation runs for up to $T = 201$ time steps, and the primary observable is the average $\log(\text{GDP}(t))$ evaluated at $t \in \{1, 11, 21, \ldots, 201\}$.

\subsection{Stylized Facts: The Role of Innovation}

\begin{figure}[h]
    \centering
    \includegraphics[width=0.85\textwidth]{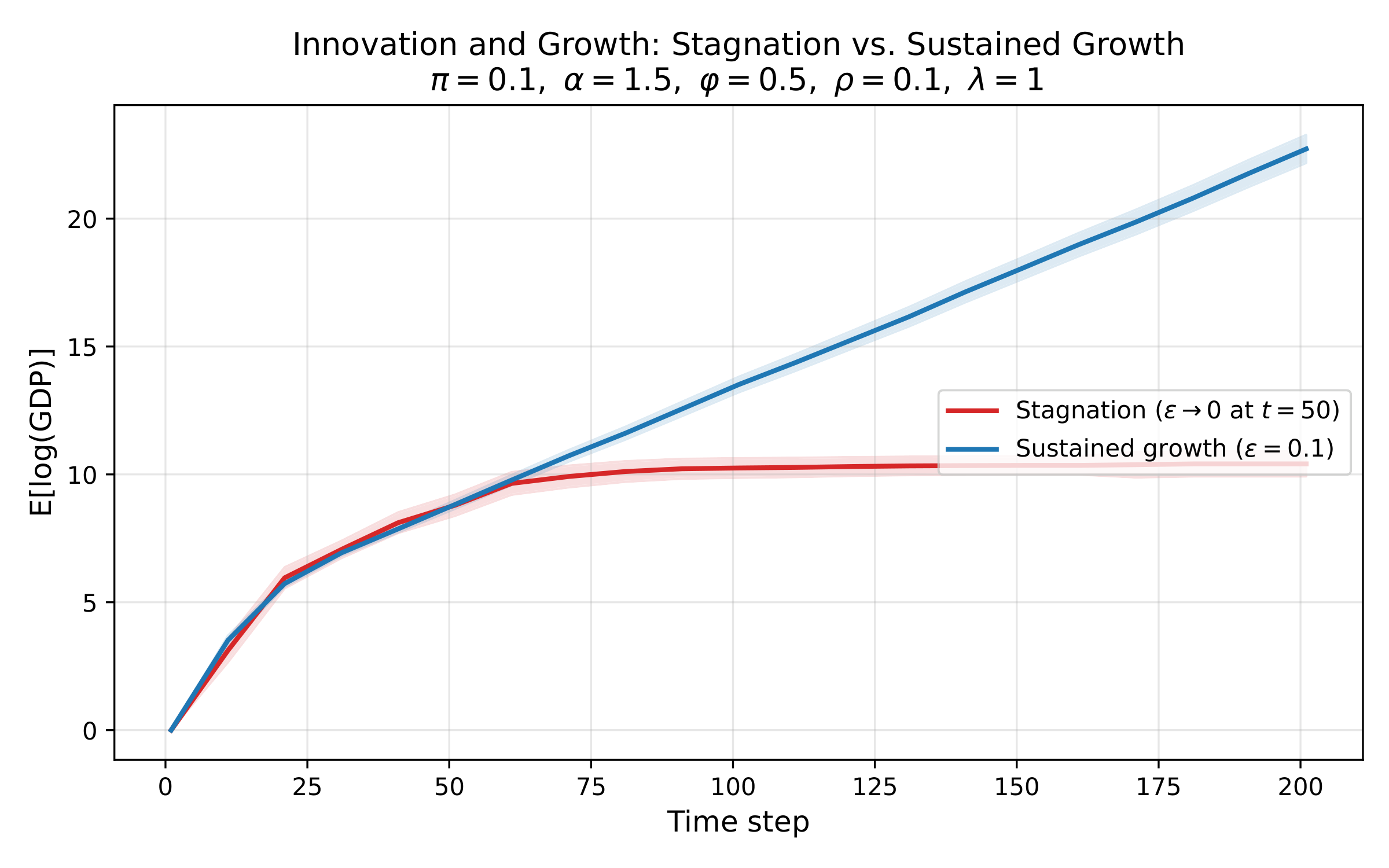}
    \caption{Stagnation vs.\ sustained growth. When exploration ceases at $t=50$, log(GDP) plateaus. With continuous exploration ($\varepsilon = 0.1$), growth is sustained. Shaded bands show 95\% confidence intervals.}
    \label{fig:stylized}
\end{figure}

A fundamental prediction of endogenous growth theory is that sustained growth requires ongoing innovation~\cite{romer1990endogenous,aghion1992model}. We verify this stylized fact by contrasting two scenarios under the baseline parameterization. This is shown in Fig.~\ref{fig:stylized}. 

In the \emph{stagnation} scenario, the exploration probability is set to zero at $t = 50$ ($\varepsilon \to 0$), removing all incentive for technological search. As shown in Fig.~\ref{fig:stylized}, average $\log(\text{GDP})$ grows during the initial phase when exploration is active, but plateaus after $t = 50$, stabilizing at around $10.4$. Without exploration, agents exhaust the productivity of known islands and no new technologies are discovered.

In the \emph{sustained growth} scenario, $\varepsilon = 0.1$ throughout. The economy exhibits approximately linear growth in log-output, reaching average $\log(\text{GDP}) \approx 22.7$ at $t = 201$---more than double the stagnation level. The two trajectories diverge sharply after $t = 50$, confirming that continuous innovation is necessary and sufficient for sustained growth. 

This result reproduces the stylized fact from Fagiolo and Dosi~\cite{fagiolo2003exploitation} (their Fig.~1a) and aligns with the core prediction of endogenous growth theory: economies that cease innovating converge to a stationary state.

\subsection{Exploration-Exploitation Trade-off}

The exploration probability $\varepsilon$ governs the fraction of miners who leave their current island to search for new technologies. While exploration drives long-run growth, it has an immediate cost: explorers do not produce output while searching, creating a trade-off analogous to the multi-armed bandit problem~\cite{sutton2018reinforcement}.

\begin{figure}[h]
    \centering
    \includegraphics[width=0.75\textwidth]{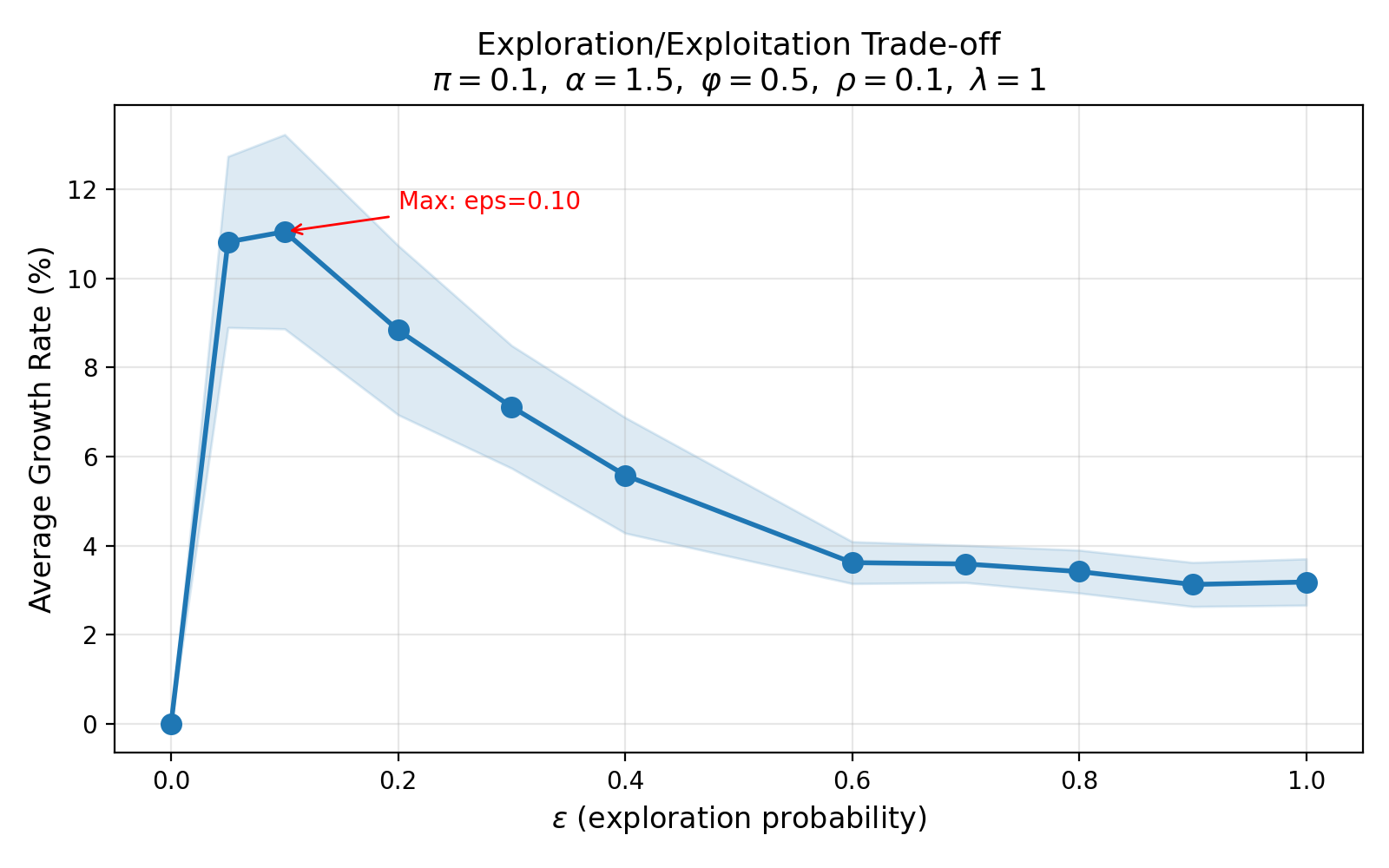}
    \caption{Average Growth Rate vs.\ exploration probability $\varepsilon$. Growth peaks at $\varepsilon \approx 0.1$, demonstrating the exploration-exploitation trade-off. Error bars show 95\% confidence intervals.}
    \label{fig:agr}
\end{figure}

Fig.~\ref{fig:agr} plots the Average Growth Rate (AGR) across 11 values of $\varepsilon \in [0, 1]$. AGR increases sharply from $\varepsilon = 0$ (no growth) to $\varepsilon \approx 0.1$, where it peaks, and then declines monotonically. At high exploration rates ($\varepsilon > 0.5$), most agents are searching rather than producing, and growth stabilizes at a lower level.

The peak at $\varepsilon = 0.1$ reproduces the finding of Fagiolo and Dosi~\cite{fagiolo2003exploitation} (their Fig.~6d): a moderate exploration rate optimally balances the discovery of new technologies against the exploitation of existing ones. This is consistent with March's~\cite{march1991exploration} theoretical prediction that organizations perform best with a balanced exploration-exploitation strategy.

\subsection{Counterfactual Analysis}

We perform a counterfactual sensitivity analysis by varying individual parameters while holding the others at their baseline values, following the methodology of Fagiolo and Dosi~\cite{fagiolo2003exploitation} (their Fig.~8). For each parameter configuration, MultiVeStA runs simulation batches (block size 30) until the confidence interval width converges below $\delta = 1$.

To formally compare the resulting trajectories across different parameter values, we use MultiVeStA's statistical hypothesis testing module, following the methodology described in~\cite{Vandin2022}. Given two sets of simulation traces obtained under different parameterizations, MultiVeStA applies a Welch's t-test~\cite{welch1947generalization} at each time step to test the null hypothesis $H_0$ that the two configurations produce equal expected values of the observable. The test computes the t-statistic from the sample means and variances of the two groups, and rejects $H_0$ at significance level $\alpha_{\text{conf}} = 0.05$ when the statistic falls outside the acceptance region. MultiVeStA also reports the statistical power of the test, quantifying the probability of correctly detecting a difference when one exists. The t-test results are shown at the bottom of each sweep figure: a filled dot ($\bullet$) indicates that the null hypothesis is not rejected (means are equal), while a cross ($\times$) indicates rejection (means are statistically different).

Out of 7 pairwise comparisons across three parameters, 6 reject the null hypothesis of equal means at $t = 201$ with power above $0.85$, confirming that the observed differences in growth trajectories are statistically significant. The only exception is $\rho = 3.0$ vs.\ $5.0$, where the test does not reject equality at any time step, suggesting that the effect of knowledge locality saturates beyond a certain threshold.

\subsubsection{Returns to Scale ($\alpha$)}

The parameter $\alpha$ controls whether production exhibits decreasing ($\alpha < 1$), constant ($\alpha = 1$), or increasing ($\alpha > 1$) returns to the number of miners on an island. We tested values $\alpha \in \{0.9, 1.0, 1.1\}$, all of which achieved convergence with $\delta = 1$. Overall, the maximum number of required simulations was $60$ for $\alpha = 1.1$ (at multiple time steps from $t = 131$ onward, due to higher variance in the super-linear regime), while $\alpha = 0.9$ and $\alpha = 1.0$ converged with $30$ simulations at most steps.

\begin{figure}[h]
    \centering
    \includegraphics[width=0.85\textwidth]{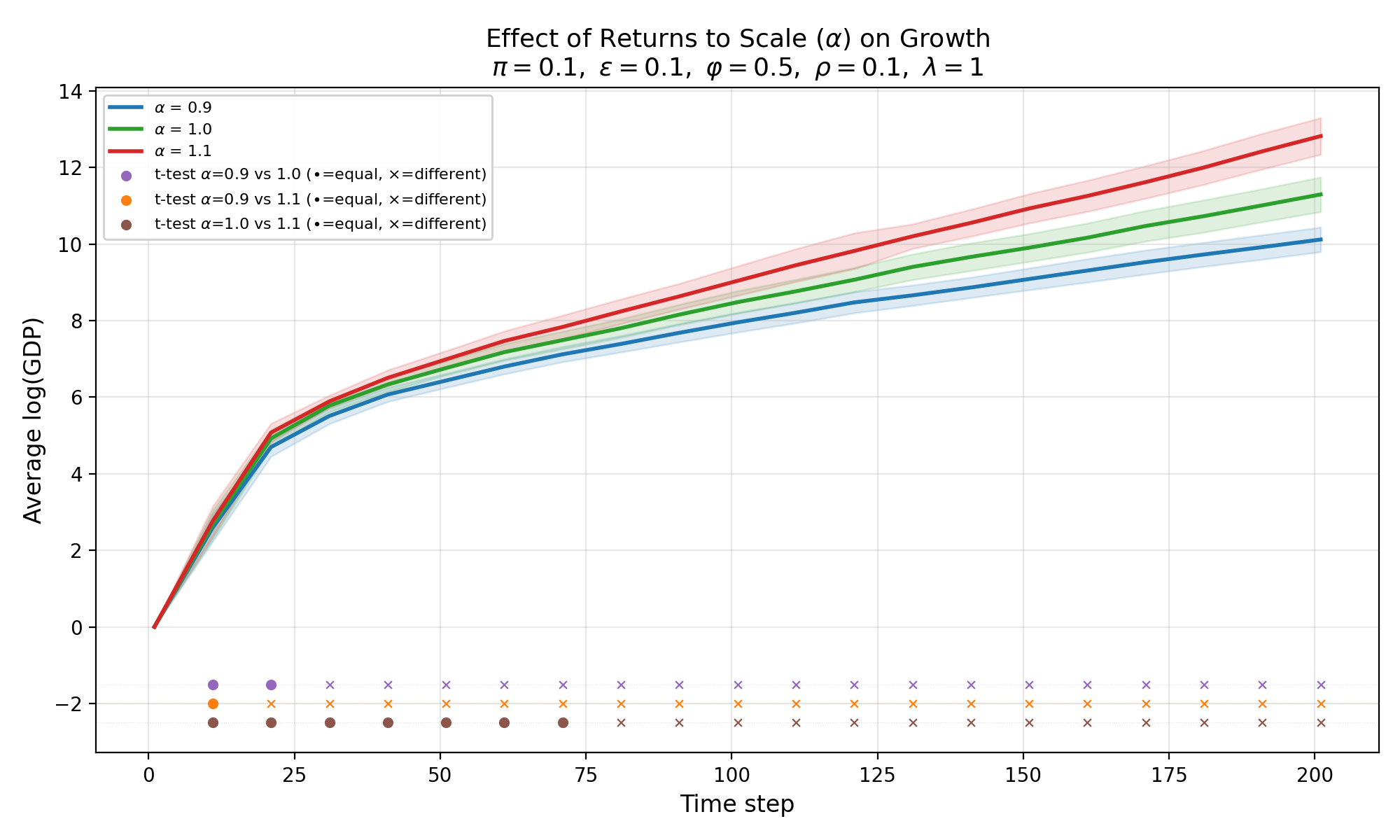}
    \caption{Effect of returns to scale $\alpha$ on average $\log(\text{GDP})$. Shaded bands show 95\% confidence intervals. Bottom: pairwise t-test results ($\bullet$ = equal means, $\times$ = different means).}
    \label{fig:alpha}
\end{figure}

Fig.~\ref{fig:alpha} shows the resulting trajectories. Growth increases monotonically with $\alpha$. Even a mild degree of increasing returns ($\alpha = 1.1$) produces markedly higher growth than constant returns ($\alpha = 1.0$), consistent with the agglomeration effects predicted by the model.

The pairwise t-tests (Fig.~\ref{fig:alpha}, bottom) confirm that all three pairs are statistically distinguishable at $t = 201$ ($\alpha_{\text{conf}} = 0.05$), with power above $0.85$ in all cases. The $\alpha = 1.0$ vs.\ $1.1$ comparison shows equal means at early time steps ($t \in [11, 71]$), with the test rejecting equality from $t = 81$ onward as the trajectories diverge.

\subsubsection{Skill Transfer ($\varphi$)}

The parameter $\varphi$ weights the contribution of an agent's past skills when determining the productivity of a newly discovered island. Only $\varphi \in \{0, 0.1\}$ achieved convergence; higher values produced variance too large for $\delta = 1$.

\begin{figure}[h]
    \centering
    \includegraphics[width=0.85\textwidth]{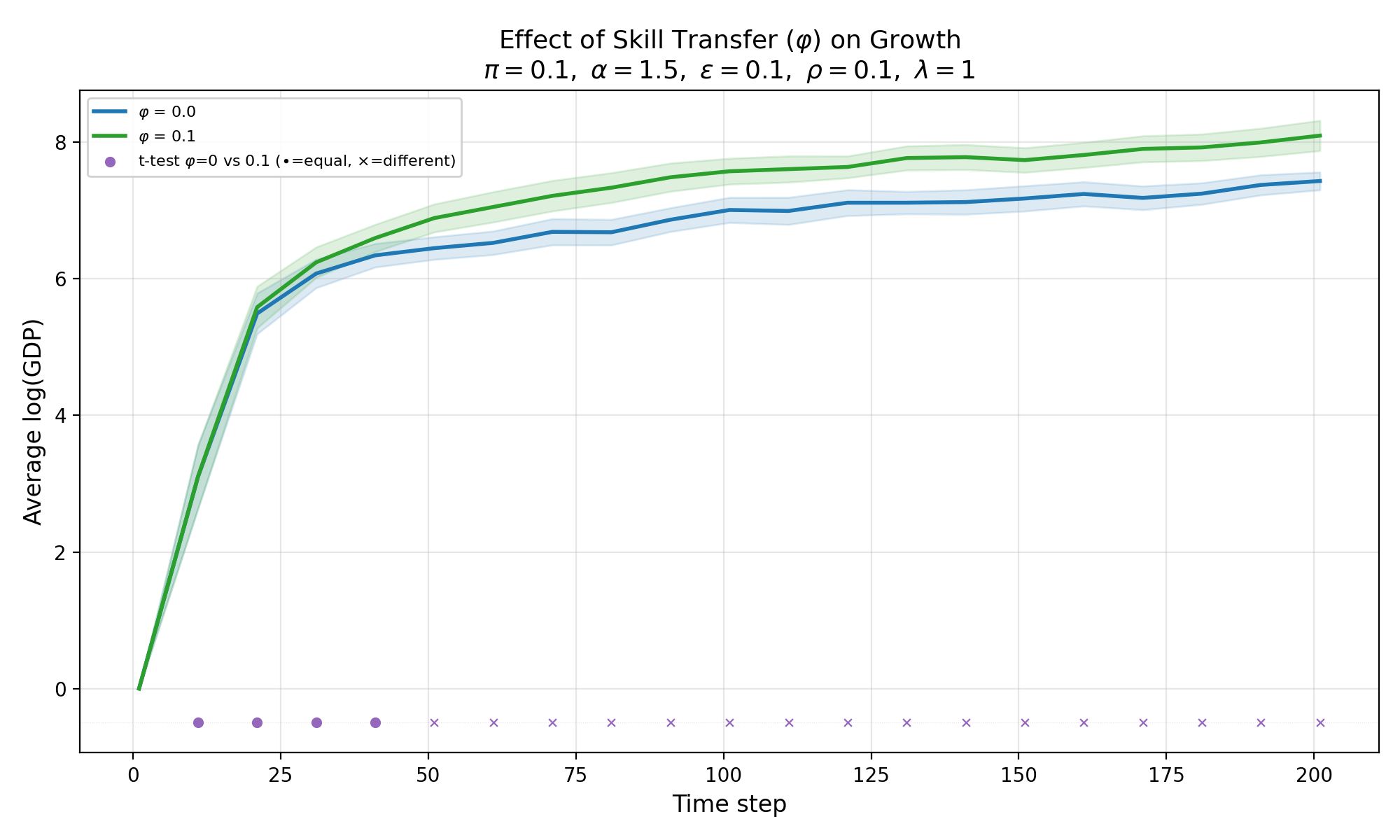}
    \caption{Effect of skill transfer $\varphi$ on average $\log(\text{GDP})$. Even low skill transfer ($\varphi = 0.1$) produces higher growth than none. Bottom: pairwise t-test results ($\bullet$ = equal means, $\times$ = different means).}
    \label{fig:phi}
\end{figure}

Fig.~\ref{fig:phi} shows that even a small amount of skill transfer ($\varphi = 0.1$) produces visibly higher growth than no transfer ($\varphi = 0$), with average $\log(\text{GDP})$ reaching $8.1$ vs.\ $7.4$ at $t = 201$. This captures the learning-by-doing mechanism emphasized by Arrow~\cite{arrow1962economic}: productivity gains are cumulative and embodied in workers' experience.

The pairwise t-test (Fig.~\ref{fig:phi}, bottom) confirms that the two means are statistically different at $t = 201$ (power $> 0.99$). At early time steps ($t \in [11, 41]$) the test does not reject equality, as the trajectories have not yet diverged sufficiently.

\subsubsection{Knowledge Locality ($\rho$)}

The parameter $\rho$ controls how rapidly productivity signals decay with distance. Low $\rho$ creates global knowledge diffusion; high $\rho$ restricts information to nearby islands. We tested $\rho \in \{1.0, 3.0, 5.0\}$.

\begin{figure}[h]
    \centering
    \includegraphics[width=0.85\textwidth]{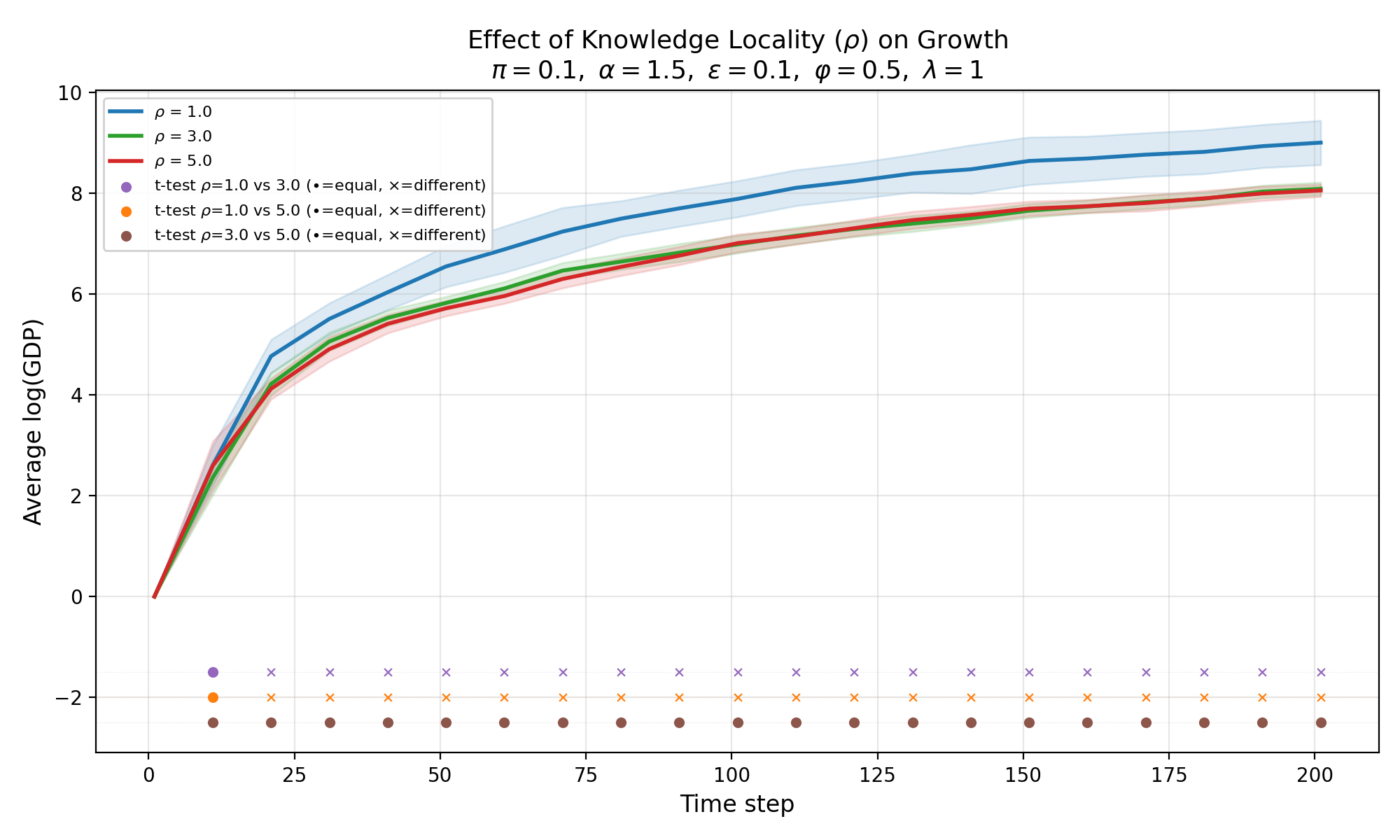}
    \caption{Effect of knowledge locality $\rho$ on average $\log(\text{GDP})$. Lower $\rho$ (broader knowledge diffusion) promotes growth. Bottom: pairwise t-test results ($\bullet$ = equal means, $\times$ = different means).}
    \label{fig:rho}
\end{figure}

Fig.~\ref{fig:rho} shows that lower $\rho$ promotes growth: average $\log(\text{GDP})$ reaches $9.0$ at $\rho = 1.0$ vs.\ $8.0$ at $\rho = 5.0$. When knowledge diffuses more broadly, agents can identify and imitate productive technologies regardless of distance.

The t-test results (Fig.~\ref{fig:rho}, bottom) reveal an interesting asymmetry: the pairs $\rho = 1.0$ vs.\ $3.0$ and $\rho = 1.0$ vs.\ $5.0$ show statistically significant differences (power $> 0.99$), while $\rho = 3.0$ vs.\ $5.0$ does not reject equality at any time step (power $= 1.0$). This suggests a non-linear relationship where the transition from broad to moderate knowledge diffusion has a measurable effect on growth, but further restricting diffusion beyond $\rho = 3.0$ does not produce additional distinguishable changes, indicating a saturation effect.

These results demonstrate that MultiVeStA's counterfactual analysis can effectively detect meaningful differences across parameter configurations. The formal statistical guarantees complement the original analysis of Fagiolo and Dosi~\cite{fagiolo2003exploitation}, confirming that variations in returns to scale, skill transfer, and knowledge locality produce genuinely distinct growth dynamics.

\section{Conclusions}
\label{sec:conclusions}
We have shown how \mv{}~\cite{sebastio2013multivesta} can automate and enrich the analysis of the Island Model~\cite{fagiolo2003exploitation}, a seminal agent-based model of endogenous growth. Our experiments reproduced key stylized facts with formal confidence intervals, confirmed the optimality of moderate exploration rates ($\varepsilon \approx 0.1$), and established through counterfactual analysis that 6 out of 7 pairwise parameter comparisons yield statistically different growth trajectories, with the single exception ($\rho = 3.0$ vs.\ $5.0$) revealing a saturation effect in knowledge locality. By automating convergence checking and hypothesis testing, \mv{} provides a principled alternative to ad-hoc Monte Carlo approaches, and a reusable template for the rigorous analysis of agent-based models across economics and beyond; the approach scales with simulation cost and variance rather than state-space size, though higher-dimensional parameter sweeps would require more selective experimental designs.

\paragraph{Future Work.}
Several directions warrant investigation. Recently, \mv{} has been integrated with process mining techniques to \emph{explain} analysis results~\cite{CasaluceBCV23,DBLP:conf/isola/CasaluceTV24,Casaluce2024,DBLP:conf/damocles/CasaluceBCLV24}; applying these to agent traces could reveal decision patterns invisible at the aggregate level, taking inspiration from techniques to discover process collaborations~\cite{DBLP:journals/sosym/CorradiniPRRT24}. We also plan to explore steady-state analyses, already supported by \mv{}, and to apply the framework to richer model variants incorporating financial sectors~\cite{Fagiolo2020InnovationFinanceGrowth} and large-scale macro-financial ABMs~\cite{dosi2010schumpeter,dosi2017micro,dawid2012eurace}, where SMC could enable rigorous policy analysis---for instance, assessing the impact of prudential regulation on long-run growth or formally testing whether financial frictions shift the economy between growth regimes. Finally, data-driven calibration against historical data and the introduction of extended agent types represent natural extensions of this work.

\bibliographystyle{eptcs}
\bibliography{generic}
\end{document}